\PassOptionsToPackage{square,sort,comma,numbers}{natbib}

\documentclass{article} 

\usepackage[preprint]{nips_2018}

\usepackage{amsmath}
\usepackage{algorithm}
\usepackage[noend]{algpseudocode}
\usepackage{microtype}
\usepackage{balance}
\usepackage{graphicx}
\usepackage{psfrag}
\usepackage{color}
\usepackage{multirow}
\usepackage{pgfplots}
\usepackage{amsthm, amsmath, amssymb, amsfonts, graphics}
\usepackage{epstopdf, psfrag, pstricks}

\title{Speaker-Independent Speech-Driven Visual Speech Synthesis using Domain-Adapted Acoustic Models}

\author{
Ahmed Hussen Abdelaziz\\
Apple Inc.\\
Cupertino, CA\\
\texttt{ahussenabdelaziz@apple.com}
\And
Barry-John Theobald\\
Apple Inc.\\
Cupertino, CA\\
\texttt{bjtheobald@apple.com}
\And
Justin Binder\\
Apple Inc.\\
Cupertino, CA\\
\texttt{jbinder@apple.com}
\And
Gabriele Fanelli\\
Apple Inc.\\
Zurich\\
\texttt{gabriele\_fanelli@apple.com}
\And
Paul Dixon\\
Apple Inc.\\
Zurich\\
\texttt{dixonp@apple.com}
\And
Nicholas Apostoloff\\
Apple Inc.\\
Cupertino, CA\\
\texttt{napostoloff@apple.com}
\And
Thibaut Weise\\
Apple Inc.\\
Cupertino, CA\\
\texttt{thibaut@apple.com}
\And
Sachin Kajareker\\
Apple Inc.\\
Cupertino, CA\\
\texttt{skajarekar@apple.com}
}

\begin{document}

\maketitle

\begin{abstract}
Speech-driven visual speech synthesis involves mapping features extracted from acoustic speech to the corresponding lip animation controls for a face model. This mapping can take many forms, but a powerful approach is to use deep neural networks (DNNs).  However, a limitation is the lack of synchronized audio, video, and depth data required to reliably train the DNNs, especially for speaker-independent models.  In this paper, we investigate adapting an automatic speech recognition (ASR) acoustic model (AM)  for the visual speech synthesis problem. We train the AM on ten thousand hours of audio-only data. The AM is then adapted to the visual speech synthesis domain using ninety hours of synchronized audio-visual speech. Using a subjective assessment test, we compared the performance of the AM-initialized DNN to one with a random initialization. The results show that viewers significantly prefer animations generated from the AM-initialized DNN than the ones generated using the randomly initialized model. We conclude that visual speech synthesis can significantly benefit from the powerful representation of speech in the ASR acoustic models.
\end{abstract} 

\vspace{2mm}

\section{Introduction} \label{sec:introduction}

Training deep neural networks (DNNs) that operate ``in the wild'' and at scale requires training on large data sets.  To avoid the requirement to train networks from scratch, a common approach is to begin with a pre-trained network, and then fine-tune the weights of this network (possibly for a different task).  An advantage of this approach to training is that the network is able to take advantage of existing knowledge, and so training is significantly faster than training from a random initialization.

In this work, we are interested in training a DNN to produce animation controls for a talking face from user speech.  In particular, we are interested in training a speaker-independent model that works for any user without the need for enrollment.   We compare two approaches:  1) training a model from scratch from a random initialization, and 2) fine-tuning an acoustic model trained for automatic speech recognition (ASR) after replacing the classification layer with a regression layer.

The remainder of this paper is organized as follows: in Section \ref{sec:related_work} we discuss related work, Section \ref{sec:methods} describes our approach for synthesizing visual speech using adapted acoustic models. The main modules of the proposed framework, which are the acoustic model and the animation control regressor, are presented in Sections \ref{sec:am} and \ref{sec:bsc}, respectively. The experiments and results of the proposed approach are described in Section \ref{sec:exp}. Finally, in Section \ref{sec:conclusions}, we summarize the paper and describe possible future work directions.

\section{Related Work} \label{sec:related_work}

Generating realistic lip motion on a face model to accompany speech is challenging because the relationship between the sounds of speech and the underlying articulator movements is complex and non-linear.  Approaches for mapping acoustic speech to facial motion fall into two broad categories:  (1) direct methods from the speech utterance itself, and (2) indirect methods, e.g.\ via phonemic transcription.  For direct approaches, the conversion function typically involves some form of regression \cite{karras2017audio,shimba2015talking,suwajanakorn2017synthesizing,taylor2016audio} or indexing a codebook of visual features using the corresponding features extracted from the acoustic speech \cite{arslan1998codebook,gutierrez2005speech}. For indirect approaches, the mapping function involves concatenation or interpolation of pre-existing data \cite{bregler1997video,cosatto2000photo,ezzat2002trainable,taylor2012dynamic,mattheyses2013comprehensive} or using a generative model \cite{anderson2013expressive,fan2015photo,kim2015decision}.
 
An advantage of mapping acoustic speech directly to visual speech is that it does not require a potentially error-prone phonemic transcription. Further, unlike phonemic features, acoustic features are rich in contextual information that corresponds to prosody and style of speech. However, learning a mapping from acoustic features to facial motion is not trivial. Firstly, it requires a sufficiently large corpus to ensure a good coverage of acoustic and visual speech for better generalization \cite{yehia1998quantitative}. Secondly, an effect of coarticulation can introduce asynchrony between the acoustic and visual modalities. In this case, a frame-to-frame mapping might be suboptimal. 

To model the temporal effects of coarticulation, many variants of hidden Markov models (HMMs) have been proposed. Inspired by the task dynamics model of articulatory phonology, one approach adopted by some text-based systems is to concatenate context-dependent phone models and sample the maximum likelihood parameters, and then use these parameters to guide the selection of samples from real data \cite{govokhina2006tda,wang2015hmm}.  Alternatively, longer phone units can be used (e.g.\ quinphones) to better capture longer-term speech (and other visual prosodic) effects \cite{anderson2013expressive}; but, these models require increasingly large training sets.  

For audio-visual models, coupled HMMs \cite{brand1999voice} can be trained, with HMM chains coupled through cross-time and cross-chain conditional probabilities \cite{xie2007coupled,abdelazizCHMM}. Coupled HMMs allow for: (1) asynchrony between the modalities and (2) the difference in effective `units' in acoustic and visual speech. To avoid the use of Viterbi decoding to compute the state sequence from which visual parameters are sampled, Baum---Welch HMM inversion was introduced in \cite{choi2001hidden,fu2005audio}.

Using HMMs to model complex multimodal signals has limitations because only a single hidden state is allowed in each time frame. This restriction means that many more states are required than would otherwise be necessary to capture the complexities of the cross-modal dynamics. To overcome this problem, dynamic Bayesian networks (DBNs) with Baum---Welch DBN inversion can be used to model the cross-model dependencies and perform the audio-to-visual conversion \cite{xie2007realistic}.

Increasingly, deep neural network (DNN) based models are used for audio-to-visual inversion.  Architectures used include fully-connect feedforward networks \cite{taylor2017deep}, recurrent neural networks (RNNs) and/or long-short term memory (LSTM) models \cite{fan2015photo,pham2017end,song2018talking,suwajanakorn2017synthesizing}, and generative adversarial networks \cite{jalalifar2018speech,vougioukas2018end}.  Many approaches are trained end-to-end to map directly from speech to video, but the approach by Taylor et al.\ \cite{taylor2017deep} achieves speaker-independence using a phonemic transcription as input. 

\section{Visual speech synthesis using adapted acoustic models}\label{sec:methods}

In ASR systems, acoustic models are responsible for estimating the posteriors of speech units given the acoustic features.  In hybrid HMM/DNN acoustic models, the speech units are referred to as senones, which are states of tri-phone HMMs.  State-of-the-art AMs typically model in the order of eight thousand senones, which provides a fine-grained representation of speech.  Our goal in this work is to determine if adapting such an ASR acoustic model, which is trained from thousands of hours of speech spoken by thousands of speakers, is better for generating animation controls than a randomly-initialized model trained specifically for the visual speech synthesis task.  We hypothesize that the discriminative power of the pre-trained acoustic model suggests that it captures a good encoding of the speech. Therefore, a fine-tuned model will be better able to predict lip motion.

\begin{figure}[t!]
\centering
\includegraphics[width=\linewidth]{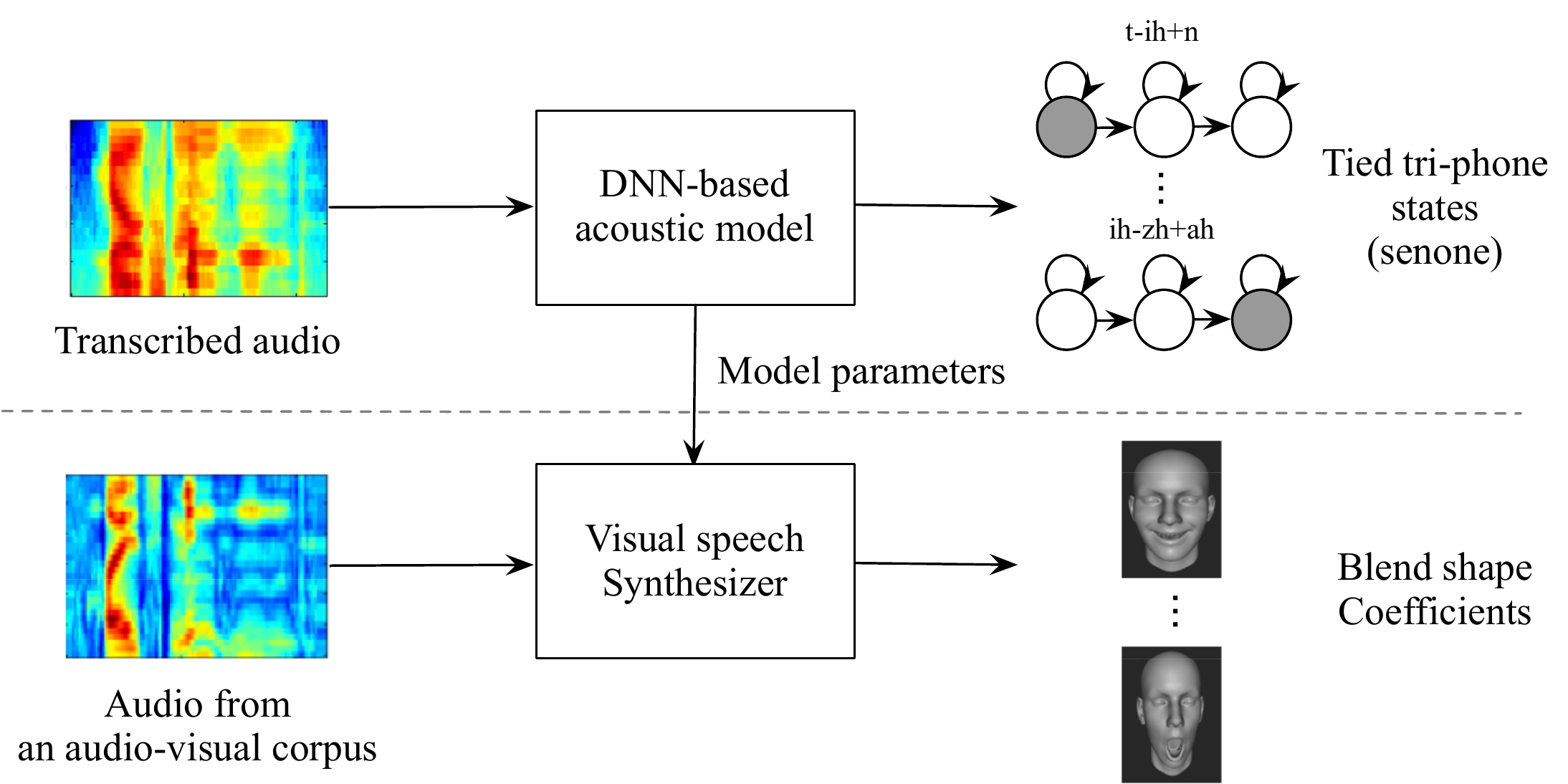}
\caption{Adapting an ASR acoustic model for visual speech synthesis. A DNN-based acoustic model is trained using acoustic features and their corresponding senones from a large corpus of transcribed speech. Model parameters of all layers prior to the softmax layer are copied to the corresponding layers of the visual speech synthesis DNN. A regression layer is added and trained on acoustic features and their corresponding animation controls (blendshape coefficients) from a smaller corpus of synced multimodal data.}
\label{fig:V_Frontend1}
\vspace{-10pt}
\end{figure}

Figure \ref{fig:V_Frontend1} shows the idea underlying our approach. First, a DNN-based acoustic model is trained to output senones using a large dataset of transcribed speech.  After convergence, the classification layer is removed, the weights of the remaining layers are frozen, and a regression layer is added to output animation controls.  As will be described in Section \ref{sec:bsc}, we use the so-called blendshape coefficients \cite{weise2011realtime} for parametrizing the face model and for controlling the animation. 

In the following subsections, we provide an outline of the  ASR acoustic model and the blendshape coefficient estimator, which are the two main modules of the proposed framework.

\subsection{Acoustic Models for ASR}\label{sec:am}

We use similar steps to the standard Kaldi receipes \cite{povey2011kaldi} for training speaker-independent DNN/HMM acoustic models. First, monophone Gaussian mixture models (GMM)/HMMs with 3 states are trained. The parameters of these models are initialized with the data global mean and variance, namely a flat start initialization procedure.  Next, triphone GMM/HMM models with 8419 tied states (senones) \cite{young1994tree} are trained. Tying the statistics of triphone HMM states is necessary to reduce the number of classes that the model needs to distinguish between. The features used for training the initial monophone and triphone models are 13-dimensional mel-frequency cepstral coefficients (MFCCs) cascaded with their first and second derivatives. The triphone GMM/HMM models are then used to estimate a linear discriminant analysis (LDA) transformation matrix. The LDA transform is applied to the raw acoustic feature vectors to create new LDA-based acoustic features. These features are then used to train new triphone models. The final triphone GMM/HMM models are speaker-adaptive-trained (SAT) models, which are trained using 40-dimensional fMLLR feature vectors. The frame-state alignments for training each triphone model are obtained by applying the forced alignment algorithm to the previously trained model. 

The last frame-senone alignments generated using the SAT models are used as the ground truth labels for training the DNN-based AM. The acoustic features used for training the DNN-based AM are 40-dimensional log-scaled, mel-scaled filter bank (LMFB) features. A window size of 21 frames centered on each frame was used to capture the phonetic context. A hamming window of 25~ms and a hop size of 10~ms were used for the LMFB feature extraction.

The DNN acoustic model has a convolutional layer with 128 filters of size $21 \times 8$, five fully-connected layers of size 1024 with SELU activation \cite{KlambauerUMH17}, a  fully connected linear bottleneck layer of size 512, and finally, a softmax output layer of size 8419.  The weights are tuned using mini-batch gradient descent to minimize the cross entropy loss. Finally, the DNN weights are re-tuned to minimize the sequential minimum bias risk (sMBR) objective function \cite{vesely2013sequence}.

\subsection{Blendshape Estimation}\label{sec:bsc}

We represent the space of facial expressions, including those caused by speech, using a low-dimensional generic blendshape model inspired by Ekman’s Facial Action Coding System~\cite{Ekman1978}. Such model can generate a mesh corresponding to a specific facial expression as:
\begin{displaymath}
\mathbf{v}(\mathbf{x}) = \mathbf{b}_0 + \mathbf{B}\mathbf{x},
\end{displaymath}
where $\mathbf{b}_0$ is the neutral face mesh, the columns of matrix $\mathbf{B}$ are additive displacements corresponding to a set of $n = 51$ blendshapes, and $\mathbf{x} \in {[0,1]}$ are the weights applied to such blendshapes.

Manually labelling the coefficients ($\mathbf{x}$) for a dataset that is large enough to train DNNs is impracticable as annotation is both subjective and time consuming. We therefore use an extension of the method described in~\cite{weise2011realtime} to automatically estimate the blendshape weights (plus the rigid motion of the head) from RGB-D video streams.

First, for each subject in our dataset, we construct a personalized model from a set of RGB-D sequences of fixed, predetermined expressions, while the head is rotating slightly. In particular, we use an extension of the example-based facial rigging method of~\cite{Li2010}, i.e., we modify a generic blendshape model to best match the specific user's example facial expressions. We improve registration accuracy over~\cite{Li2010} by adding positional constraints to a set of 2D landmarks detected around the main facial features in the RGB images using a CNN similarly to~\cite{He2017AFE}.

Next, given a personalized model, we automatically generate labels for the head motion and blendshape coefficients (BSCs) by tracking every video frame of the same subject. We first rigidly align the model to the depth maps using iterative closest point (ICP) with point-plane constraints. We then solve for the blendshape coefficients which modify the model non-rigidly to best explain the input data. In particular, we use a point-to-plane fitting term on the depth map:
\begin{displaymath}
{D_i}(\mathbf{x}) ={  \left( {\mathbf{n}_i}^\top  \left( \mathbf{v}_i(\mathbf{x}) - \overline{\mathbf{v}}_i \right)  \right) }^{2},
\end{displaymath}
where $\mathbf{v}_i$ is the i-th vertex of the mesh, $\overline{\mathbf{v}}_i$ is the projection of $\mathbf{v}_i$ to the depth map, and $\mathbf{n}_i$ is the surface normal of $\overline{\mathbf{v}}_i$.  Additionally, we use a set of point-to-point fitting terms on the detected 2D facial landmarks:
\begin{displaymath}
{L_j}(\mathbf{x}) = {{|| \pi ( \mathbf{v}_j(\mathbf{x})   ) - \mathbf{u}_j  ||}}^{2},
\end{displaymath}
where $\mathbf{u}_j$ is the position of a detected landmark and $ \pi ( \mathbf{v}_j(\mathbf{x}) )$ is its corresponding mesh vertices projected into camera space. The terms $D_i$ and $L_j$ are combined in the following optimization problem:
\begin{displaymath}
\underset{ \mathbf{x} }  {\text{min}} \  w_d \sum_i D_i ( \mathbf{x} ) + w_l \sum_j F_j ( \mathbf{x} ) + w_{r} || \mathbf{x} ||_1,
\end{displaymath}
where $w_d$, $w_l$, and $w_r$ represent the weights given respectively to the depth term, the landmark term, and a L1 regularization term, which promotes the solution to be sparse. The minimization is carried out using a solver based on the Gauss-Seidel method.

\section{Experimental Results} \label{sec:exp}

\subsection{Datasets}

For training the acoustic model, we used an in-house corpus that contains around ten thousand hours of query utterances for a digital assistant. A separate dataset of 9 hours, almost seven thousand utterances, was used to evaluate the model. 

The data used for training the visual speech synthesis model contains around ninety hours of multimodal data. Each utterance has 44.1kHz, 16bps PCM audio, 60~frame per second (fps) RGB, and 30~fps depth streams. Subjects are balanced demographically. The corpus is divided into 75/11/4 hours for training, validation, and testing respectively. 

\subsection{Experimental setup}\label{sec:setp}

As described in Section \ref{sec:am}, the DNN-based AM is trained using LMFB acoustic features. Both the input LMFB features and the output senone posteriors are estimated at a frame rate of 100~fps. On the other hand, the BSCs described in Section \ref{sec:bsc} are produced at the video frame rate of 60~fps. To match these frame rates, the ground-truth senones are down-sampled to 60~fps and a new set of acoustic features are extracted with a hop size of 16.67~ms. With that hop size, the window used for extracting the LMFB features spans almost 350~ms of audio compared to 210~ms using the original 10~ms hop size. The impact of the hop size on the AM performance is discussed in more detail in Section \ref{sec:results}. The DNN-based AM is retrained using the new features and the down-sampled ground-truth senones to minimize the sMBR objective function.

After training the AM, all parameters of all layers prior to the bottleneck layer are frozen and the \texttt{softmax} layer of the AM is replaced by a 32-dimensional regression layer. The 32 components of the output vector are the BSCs that control the lips, the jaw, the cheek, and the mouth of the 3D face model. The parameters of the regression layer, which is a linear transform from the acoustic bottleneck features to the BSCs, are estimated using mini-batch gradient decent and back propagation so that the mean absolute error (MAE) between the ground truth BSCs and the network inference is minimized. The ninety hours of the multimodal data are used for tuning the \emph{final} regression layer parameters.  

Before rendering, the output of the network is post-processed as follows:  a moving median of length 5 is used to smooth the output to avoid unnatural abrupt changes. Another moving window of 60 frames is used for bias normalization. Within this window, the minima of the BSCs are estimated and averaged. This average is then subtracted from the BSCs. Finally, a global scale of 1.5 (determined empirically) is applied to boost the articulation. Another scaling factor of 2.5 is applied to the lip pucker and lip funnel coefficients (also determined empirically). 

\subsection{Results} \label{sec:results}

Table \ref{table:AM_FR} compares the frame accuracy of DNN-based AMs on the test set when trained with 60~fps and 100~fps acoustic features using both the cross entropy and the sMBR loss functions.  For both losses, the 60~fps model outperformed the 100~fps model.  This is likely because  the same utterances are used for training and testing the models, so models trained with higher frame rate features have more data ($\approx 1.9$~million for the 60~fps and $\approx 2.9$~million for the 100~fps).  Further, the context window used for acoustic feature extraction spans more data in the case of 60~fps than the 100fps acoustic features. 

As expected, Table \ref{table:AM_FR} shows that models trained using the sMBR loss function achieve slightly better accuracy than those trained using the  cross entropy loss. In general, the accuracies shown in Table \ref{table:AM_FR} are in the typical accuracy range of our in-house state-of-the-art, production quality ASR acoustic models. These results ensure the reliability of the AM after the feature/label downsampling process and validates the use of the 60~fps AM in the next steps.

\begin{table}[t!]
	\centering{
		\caption{Frame accuracy of AMs trained using 100~fps and 60~fps acoustic features and the cross entropy (CE) and the sequential minimum bias risk (sMBR) loss functions.}
		\label{table:AM_FR}
			\begin{tabular}{ccc}
				\hline
				\multirow{2}{*}{Frame rate} 	&   \multicolumn{2}{c}{ Eval set frame accuracy [\%]} \\
				 											&  CE loss 		& sMBR \\
				\hline                                                                   
				100~fps  								& 66.8\% 				& 67.1\%   \\
				60~fps    							    & 69.5\%   			&  70.1\%  \\
			\hline                                                           
		\end{tabular}
	}
	\vspace{-12pt}
\end{table}

Table \ref{table:VSSM_MAE} shows the mean absolute error (MAE) between the ground truth BSCs and the network inference in the test set. As shown, the difference between the two initialization methods is not conclusive. Despite being objectively similar, the rendered videos from the two models are perceptually different. The reason for this is that visual speech can be well inferred from the acoustic features, but other facial movements, such as smiles, cannot. The contribution to the error from the speech-related BSCs is small compared to the contribution from the other BSCs. Thus, we use subjective testing to quantify the quality of the models.

\begin{table}[t!]
	\centering{
		\caption{Mean absolute error of the visual speech synthesis models (see Figure \ref{fig:subjective} for subjective measures).}
		\label{table:VSSM_MAE}
			\begin{tabular}{cc}
				\hline
				Initialization method 	&  MAE of the test set  \\
				\hline                                                                   
				Initialization with AM					    & 0.0857  \\
				Random initialization    					& 0.0855  \\
			\hline                                                           
		\end{tabular}
	}
\vspace{-12pt}
\end{table}

\begin{figure}[t!]
\centering
\includegraphics[width=0.5\linewidth]{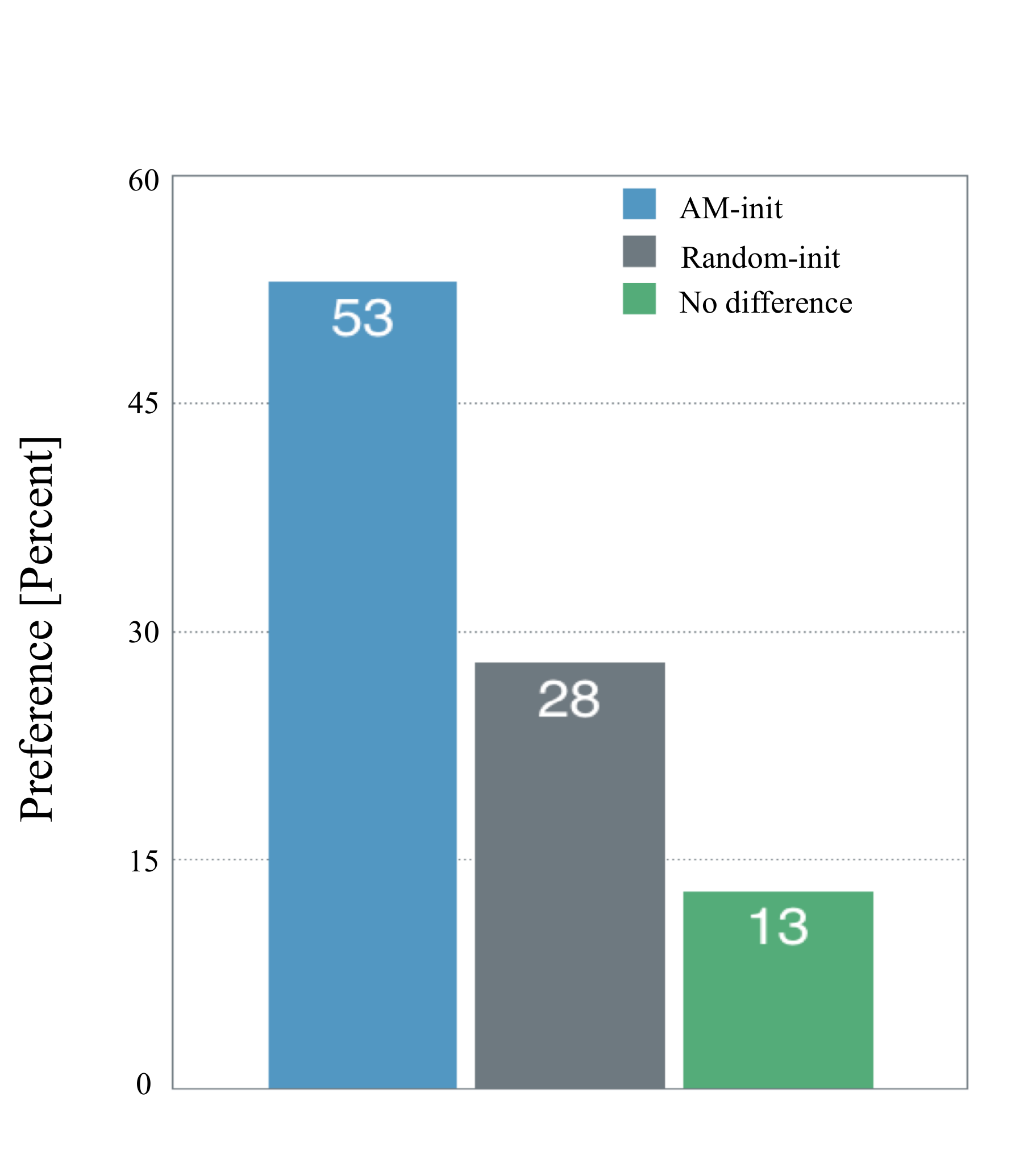}
\caption{Subjective assessment test results for comparing the quality of speaker-independent visual speech synthesis.}
\label{fig:subjective}
\vspace{-12pt}
\end{figure}

To perceptually compare the output of the two models, we conducted a subjective test where viewers were presented with a pair of sequences, and they were asked  ``which video matches the speech more naturally''.  In total, 30 utterances were selected at random. The corresponding videos were generated using networks that were trained using the random and the acoustic model initialization. To prevent display ordering effects, the order in the pair that the videos were presented was randomized.  In total, 30 graders evaluated videos for 30 utterances. Figure \ref{fig:subjective} shows the percentages for viewers preferring each of the approaches (including no difference).  As shown, the AM-initialized network generates more natural visual speech than the randomly initialized network. A Mann-Whitney test \cite{mann1947} shows that the results are statistically significant ($u=0.65, p < 0.0001$). 

In addition to selecting their preference for a sequence, graders were asked to justify their preference. Graders commented that the AM-initialized network produced more natural lip motion, which was more in sync with the audio, than the equivalent lip motion from the randomly-initialized network. Some graders preferred the AM-initialized model for the responsiveness of lip movements to vocal effort. Also, pursed lip movements were more natural with the AM-based network. Finally, graders observed better pause formation with the AM-initialized network where lips close appropriately.

\section{Conclusions and Further Work} \label{sec:conclusions}
In this study, a speech-driven speaker-independent visual speech synthesis system was introduced. The system uses a deep neural network (DNN) that is trained on thousands of hours to extract abstract speech-related features. A regression layer is then added to predict the blendshape animation controls. A subjective assessment was conducted to compare the proposed approach to a baseline approach using a random initialization of the DNN. The subjective test shows that animations produced by the proposed system are significantly superior to the baseline.  Because the regular evaluation loss used in training the DNNs, e.g., mean absolute error, was shown to not necessarily reflect the actual performance, our future work will include the development of an objective measure to predict subjective opinion of animation quality.

\section{Acknowledgments}
The authors are grateful to Russ Webb, Stephen Young, and Shivesh Ranjan for their valuable comments.

\bibliographystyle{plain}
\bibliography{bibliography}

\end{document}